\documentclass[aps,preprint]{revtex4}
\usepackage{graphicx,amsmath,amssymb}

\newcommand{\bea}{\begin{eqnarray}}
\newcommand{\eea}{\end{eqnarray}}
\newcommand{\be}{\begin{equation}}
\newcommand{\ee}{\end{equation}}
\newcommand{\lbl}{\label}

\begin{document}

\begin{titlepage}

\title{Linking density functional and mode coupling models for supercooled liquids}

\author{Leishangthem Premkumar, Neeta Bidhoodi, Shankar P. Das}
\affiliation{School of Physical Sciences,\\
Jawaharlal Nehru University,\\
New Delhi 110067, India.}

\setcounter{equation}{0}

\begin{abstract}

We compare predictions from two familiar models of the metastable
supercooled liquid respectively constructed with thermodynamic and
dynamic approach. In the so called density functional theory (DFT)
the free energy $F[\rho]$ of the liquid is a functional of the
inhomogeneous density $\rho({\bf r})$. The metastable state is
identified as a local minimum of $F[\rho]$. The sharp density
profile characterizing $\rho({\bf r})$ is identified as a single
particle oscillator, whose frequency is obtained from the parameters
of the optimum density function. On the other hand, a dynamic
approach to supercooled liquids is taken in the mode coupling theory
(MCT) which predict a sharp ergodicity-nonergodicity transition at a
critical density. The single particle dynamics in the non-ergodic
state, treated approximately, represents a propagating mode whose
characteristic frequency is computed from the corresponding memory
function of the MCT. The mass localization parameters in the above
two models (treated in their simplest forms) are obtained
respectively in terms of the corresponding natural frequencies
depicted and are shown to have comparable magnitudes.
\end{abstract}

\vspace*{.8cm}


\maketitle
\end{titlepage}

\section{Introduction}

The dynamics of a liquid supercooled below its freezing point slows
down drastically. The relaxation time of the liquid increases
sharply with supercooling so that below a temperature, generally
termed as a calorimetric glass transition temperature $T_g$, the
liquid stops flowing and behaves like a solid with elastic
properties. The amorphous solid is in a metastable equilibrium state
with its constituent particles vibrating around their respective
parent sites which  form a disordered lattice without any long range
order. Over time scales of structural relaxation such a description
persists. The issue of an underlying thermodynamic transition in the
supercooled liquid accompanying this vitrification process in to a
solid has been studied in different models \cite{rfot}. For
describing the amorphous solid like state both thermodynamics and
dynamics based models have been proposed.

We first outline the thermodynamic approach. The transition of the
dense liquid into a crystalline state with long range order occur at
the freezing point. This transition is of thermodynamic origin. An
order parameter based model for understanding the freezing
transition and the inhomogeneous crystalline state is provided in
the density functional theory (DFT) \cite{evans,ys-pr,lowen-pr}. The
density functional approach to describe the freezing
phenomena\cite{tvr,oxtoby1}, has a statistical mechanical basis.
This method has also been extended to describe the supercooled
liquid state. Metastable minima of the free energy functional
intermediate between the crystal and the liquid, have been observed
by several authors\cite{ysingh,cdg,lowen,prl1} in the past. The
amorphous state is characterized with an inhomogeneous density
$\rho({\bf r})$ having peaks centered around a set of points $\{{\bf
R}_i\}$, constituting  an amorphous lattice without any long range
order. In DFT, sharply localized peaks of the density profiles are
identified as representing individual particles vibrating around a
lattice site. The width of the density profile signifies the average
displacement of the particle around the lattice site. This width is
required to remain within a limit\cite{lindeman} in order for the
solid like state to survive. Using the average particle displacement
from the DFT to the average potential energy, the spring constant
and hence the natural frequency for the oscillator presenting the
single particle dynamics is obtained.

Next we consider the model based on dynamics of the liquid. The mode
coupling theory (MCT) is a formulation extensively developed over
last thirty years for understanding slow relaxation in a strongly
correlated liquid. Using a microscopic approach, MCT has provided a
model for understanding the slow dynamics at the initial stage of
supercooling of the metastable liquid
\cite{my_rmp,kawa_trans,reichman}. The dynamics of the fluid is
described in terms of that of the slow modes which reflects the
conservation laws in the fluid. In the deeply supercooled state,
solid like behavior with finite shear modulus develops
\cite{dlength} and manifests through existence of transverse sound
modes characteristic of the solids. From a theoretical level this is
understood in terms of Goldstone modes which appear in the glassy
states \cite{yeomaj,rolf} and has been used to explain
characteristic features \cite{bpeak-pre,bpeak-pla} of the amorphous
states. The MCT in its simplest form predicts a transition from the
ergodic liquid state to a non-ergodic glassy state. The latter is
characterized by a finite shear modulus. The MCT is based on the
equations of microscopic dynamics\cite{spd-jsp} of a set of fluid
particles. The equations of fluctuating nonlinear hydrodynamics
(FNH) which give rise to the basic equations of the MCT, involve a
driving free energy functional $F[\psi]$, expressed in terms of the
slow modes $\{\psi\}$ of the fluid. The free energy functional for
the liquid is dependent on the density function $\rho({\bf r})$.
Thus equations of FNH are obtained using the same free energy
functional as that of standard DFT. The stationary state for the
fluid is described by the solution $\exp [-F(\psi)]$. The form of
the memory function \cite{boon-yip} giving rise to the oscillatory
behavior of the single particle correlation is obtained from these
equations \cite{bin2}.

Kirkpartick and Wolyness first demonstrated \cite{dftmct} the link
between the two basic models of the metastable liquids and obtained
the mass localization parameter in amorphous state. In the
simplified model these authors used the so called vineyard
approximation to link the collective and single particle density
correlations. It was shown that the spring constant for the
vibrating particle is obtained using respectively the description
based on static and dynamic considerations. From a static or
thermodynamic point of view the spring constant can be inferred from
the equilibrium density distribution in the inhomogeneous state. On
the other hand in a dynamical approach, the spring constant is
obtained from the frequency of the propagating oscillatory mode in
the non-ergodic state. The spring constant for vibrating particle is
computed from the long time limit of the corresponding memory
function of the MCT. In the present work we explore this link on the
DFT and MCT descriptions of the amorphous state. We compare the two
models by explicit computation of the corresponding localization
parameters with a DFT and MCT approach. The structure function of
the uniform liquid state and assumption of an amorphous lattice are
the only required inputs in the comparison between the two models.
The paper is organized as follows. In the next section we discuss
briefly the formulation of the density functional model and show how
the optimum value of the mass localization parameter follows from
purely thermodynamic considerations. Next, in section III we discuss
the dynamic model demonstrating how the mode coupling approach is
used to obtain the natural frequency of the single particle mode in
the metastable liquid. In section IV, we discuss how the MCT and the
DFT are linked . In section V we present numerical results focusing
on the key quantity for comparing the two models. We end the paper
with a brief discussion of the results.

\section{The Thermodynamic model}

In the thermodynamic approach, a generalized free energy functional
reaches a local minimum corresponding to the metastable glassy
state. The free energy functional is treated as a function of the
density which depict the inhomogeneous mass distribution in the
supercooled liquid. We present below briefly the calculation of the
free energy.

\subsection{Density functional theory}

Using the thermodynamic extremum principle for the free energy
functional $F[\rho({\bf r})]$, the inhomogeneous density $\rho({\bf
r})$ for the equilibrium state is obtained. The total free energy is
 a sum of two parts, the ideal gas term and the
interaction or the so called excess contribution, respectively
denoted as $F_\mathrm{id}$ and $F_\mathrm{ex}$:
\be F[ \rho({\bf r})] = F_\mathrm{id} [ \rho({\bf r})] +
F_\mathrm{ex}[ \rho({\bf r})]. \ee
The ideal gas part of the free energy functional is generalized for
non-uniform density $\rho({\bf r})$ as
\be \label{fid_ge} F_\mathrm{id} [\rho({\bf r})] = k_B T\int d{\bf
r} \rho({\bf r})\left(\ln[\wedge^{3} \rho({\bf r})]-1\right). \ee
The excess part is also obtained as an expansion in terms of the
density fluctuations. The density function $\rho({\bf r})$ signifies
the nature of the mass distribution in the inhomogeneous state. The
free energy $F[\rho]$ of the solid state is obtained in terms of the
density fluctuations, by using a functional Taylor expansion the
around the uniform liquid state. We consider perturbation expansion
of the excess free energy around an uniform system  of density same
as the average density of the corresponding inhomogeneous structure.
\be \rho_0=\frac{1}{V}\int d {\bf r} \rho({\bf r})~~. \ee
The functional Taylor series expansion about the homogeneous density
distribution $\rho_0$ is given as \cite{tvr},
\bea \label{fre2} \beta{F}[\rho({\bf r})]- \beta{F}(\rho_0) &=& \int
d {\bf r} \rho({\bf r})~
\ln \left( \frac{\rho({\bf r})}{\rho_0} \right) \\
\nonumber & &-\frac{1}{2}\int d {\bf r}_1 \int d {\bf r}_2 \delta
\rho({\bf r}_1) \delta \rho({\bf r}_2) c({\bf r}_1-{\bf
r}_2;\rho_0)~~, \eea
where $c(r;\rho_0)$ is the Ornstein-Zernike two point direct
correlation function \cite{hansen} for the uniform liquid of density
$\rho_0$. Here $\delta \rho({\bf r})$ is the deviation of the
inhomogeneous state density $\rho({\bf r})$ from the average density
$\rho_0$.
The free energy functional $F[\rho]$ is minimized with respect to an
optimum choice for the inhomogeneous density function $\rho({\bf
r})$. The density function is chosen a parametric form signifying
localized density profiles centered on a set of lattice points for
which we need a specific underlying lattice $\{ {\bf R}_i\}$ as an
input.

%%%%%%%  Density %%%%%%%%%

\subsection{The Oscillator model} \label{sec1.2}

In this the metastable glassy state is characterized as one in which
the individual particles are oscillating around a set of lattice
sites $\{ {\bf R}_i\}$ situated on a random lattice. The degree of
mass localization in the metastable state of the supercooled liquid
is identified from $\rho({\bf r})$ for which the free energy
$F[\rho]$ reaches a local minimum. The test density function which
has been most effectively used \cite{tarazona} in the DFT is in
terms of normalized gaussian functions centered around a set of
lattice points $\{ {\bf R}_i\}$.
\begin{equation}
\label{tar-dens} \rho({\bf r}) = {\left (\frac{\alpha}{\pi}\right
)}^{3/2} \sum_{{\bf R}_i} e^{-\alpha({\bf r}-{\bf R}_i)^2} \equiv
\sum_{i=1}^N \phi_\alpha (|{\bf r}-{\bf R}_i|)
\end{equation}
where $\phi_\alpha (r)= {(\alpha/\pi)}^{3/2} \exp ( -\alpha r^2)$.
The density is parameterized in terms of the mass localization
parameter $\alpha$ which is inversely proportional to the square of
the width of Gaussian profiles in the expression on the right hand
side of Eqn. (\ref{tar-dens}). The parameter $\alpha$ quantifies the
motion of particles in the system in a coarse grained manner. The
$\alpha \rightarrow 0$ limit depicts gaussian profiles of infinite
width and thus the corresponding phase represents the homogeneous
liquid state. Increasing values of $\alpha$ represent increasingly
localized structures and thus referring to greater inhomogeneity in
the system. The $\alpha$ corresponding to the free energy minima
determines the preferred thermodynamic phase. For simplicity we
assume that the $\alpha$ values at the different sites are the same
though in reality they are different characteristic of the
heterogeneous state \cite{chandan,prl2}.

In case of completely non overlapping Gaussian profiles, the
situation corresponds to individual particles oscillating around
their mean positions which form an underlying lattice. In case of a
crystal this lattice has long range order while for an amorphous
solid this is a random structure representing a metastable liquid.
In the harmonic solid the spring constant $\kappa_s$, for the
oscillation of a single particle around the lattice site, is simply
related to the width parameter $\alpha$ for the gaussian profiles.
The average kinetic and potential energies of an oscillator with
position and momentum coordinates $\{x,p\}$ are same and each is
equal by the law of equipartition to $(k_BT)/2$, {\em i.e.},
\begin{equation}\label{eqpart-trk} \frac{<p^2>}{2m}=
\frac{\kappa_{s}}{2}<x^2> = \frac{k_BT}{2}~~.
\end{equation}
On the other hand, corresponding to the DFT expression for the
density (\ref{tar-dens}) the average mean square displacement for
the single particle is given by, $<x^2>=1/(2\alpha)$. Using this in
the above relation we obtain
\be \label{alreln1} \alpha=\frac{\kappa_s}{2k_BT} \ee
Therefore the characteristic frequency $\omega_0$ of the oscillator
is obtained as
\be \label{alreln2} \omega_0^2=\frac{\kappa_s}{m}=2{v_0^2}\alpha~~,
\ee
where $v_0^2=k_BT/m$ is the thermal speed.

For the metastable harmonic solid held at a finite temperature the
width parameter for the inhomogeneous density function is obtained
by invoking the standard thermodynamic extremum principles. The free
energy of the metastable liquid is obtained by evaluating the
corresponding density functional expression with a test density
function for inhomogeneous state. The latter is parameterized in
terms of the width parameter $\alpha$. The optimum value of
$\alpha{\equiv}\alpha_\mathrm{DFT}$(say) corresponding to which the
free energy is a minimum presents the metastable state. Since the
particle positions are strictly localized in this case, the width
parameter ${\alpha\sigma^2}>>1$, where $\sigma$ is a microscopic
length scale associated with the interaction potential of the liquid
particles.

\section{The dynamic approach}
\label{sec2}

The self-consistent mode coupling theory (MCT) presents a
microscopic model for understanding the slow relaxation behavior
seen in a supercooled liquid using a dynamic approach. The dynamics
of the dense liquid is generally described in terms of that of a set
of slow modes which signify the conservation laws for the fluid. The
equations of motions for these modes are the balance equations
representing the underlying conservation laws. The MCT is formulated
by taking in to account the effects of the nonlinearities present in
these equations on the transport properties of the liquid. We
briefly outline below how the equations of generalized hydrodynamics
give rise to the MCT equations for glassy dynamics.

\subsection{Generalized Hydrodynamics}

%% Basic equations of FNH

The generalized Langevin equation, leads to the equations of motion
for the respective coarse grained densities of mass and momentum as
$\psi_i\equiv\{\rho({\bf r},t),{\bf g}({\bf r},t)\}$ for a one
component liquid. Following standard procedures\cite{cup} we obtain:
\bea && \label{cont-eqn}
\frac{\partial\rho}{\partial t}+\nabla . {\bf g}=0, \\
&& \label{momt-eqn} \frac{\partial {\rm g}_i}{\partial t}+ {\bf
\nabla} \cdot ( g_i{\bf v})+\rho\nabla_{i} \frac{\delta
F_{U}}{\delta\rho} - L^0_{ij}{{\rm v}_j} =\theta_{i}~~. \eea
In the following we use the convention that repeated indices are
summed over. The field ${\bf v}({\bf r},t)$ is defined in terms of
the nonlinear constraint ${\bf g}=\rho{\bf v}$ and represents
$1/\rho$ nonlinearity in the equations fluctuating nonlinear
hydrodynamics. $L_{ij}^0$ represents the matrix of bare or short
time viscosities. These dissipative coefficients are related to the
correlation of the gaussian noises respectively in Eqns.
(\ref{momt-eqn}).
\be \label{cnoise-cor} \left\langle \theta_i({\bf r},t)\theta_j({\bf
r}^{\prime},t^{\prime}) \right\rangle =
2{\beta^{-1}}L^0_{ij}\nabla^2 \delta({\bf r}-{\bf
r}^{\prime})\delta(t-t^{\prime}) ~~~. \ee
Here $\beta^{-1}$ denotes the Boltzmann factor determining the
strength of the thermal noise correlations. For an isotropic system
the bare viscosity matrix $L_{ij}^0$ involves two independent
coefficients.
\be \label{visc-tensor} L^0_{ij}= \Gamma_0\nabla_i\nabla_j
+\varsigma_0 (\delta_{ij}{\nabla}^2-\nabla_i\nabla_j)\ee
where $\Gamma_0$ and $\varsigma_0$ respectively denotes the bare or
short time longitudinal and shear viscosities. The stationary
solution of the Fokker-Planck equation corresponding to the above
stochastic equations is $e^{-F}$ with $F[\psi]$ being identified as
the free energy functional of the local densities $\{\psi({\bf
r},t)\}$. $F$ is expressed as
\be F[\rho,{\bf g}]=F_K[\rho,{\bf g}]+F_U[\rho], \ee
where the kinetic part dependent on momentum density ${\bf g}$ is
obtained as \cite{lturski}
\be F_K=\int d{\bf r} \frac{g^2({\bf r})}{2\rho({\bf r})}~~. \ee
The $1/\rho$ nonlinearity in the expression for $F_K$ is important
in producing the form of the equations of fluctuating nonlinear
hydrodynamics (FNH) for the set $\{\rho,{\bf g}\}$. The so called
potential part $F_U[\rho]$ which is a functional of density only and
is identified with the expression (\ref{fre2}) of the free energy
functional $F[\rho]$ in terms of the inhomogeneous density function
$\rho({\bf r})$ used in the DFT. Using the leading order correction
to the transport coefficients in the renormalized correlation
function, the simplest form of the mode coupling model follows. This
model predict a sharp ergodicity-nonergodicity (ENE) transition at a
critical density arising from a nonlinear feedback mechanism due to
coupling of dominant density fluctuations. The transition is defined
in terms of the long time limit of the normalized collective density
correlation function $\phi(q,t)$.
\be \label{col-den} \phi(q,t\rightarrow{\infty}) = f(q)~~. \ee
$f(q)$ is referred to as the corresponding non-ergodicity parameter
at wave number $q$ for the collective density correlation function.
As the transition is approached from the liquid side, $f(q)$ is
nonzero beyond the ENE transition. In the non-ergodic state $f(q)$
remains nonzero and shows interesting scaling behavior
\cite{gtze84,jcp-scaling} in the vicinity of the ENE transition
point. The mechanism for slow dynamics is understood in terms of the
Laplace transform of the correlation functions. The Laplace
transform of $\phi(q,t)$ is given by
\be \lbl{corre} \phi(q,z)=\left[z-\frac{q^2c_0^2}{z+i
q^2{L}(q,z)}\right]^{-1} \ee
where $c_0$ is the speed of sound and ${L}(q,z)$ is the generalized
transport coefficient or the memory function. The generalized
longitudinal viscosity $L(q, z)$ contains the bare and the mode
coupling part, expressed as
\be L(q, z) = L_{B}(q) +L^{mc} (q, z) \ee
The uncorrelated collisions occurring during the short time are
responsible for the bare contribution $L_{B}(q)$ to the viscosity.
The inverse Laplace transform of the mode-coupling contribution
$L^{mc}(q, z)$ in equation (\ref{corre}) is given by
\bea \label{visc-qdep} L^{mc}(q,t)&=&\frac{\beta^{-1}}{2 \rho_{0}}
\int\frac{d{\bf k}}{(2\pi)^{3}} V({\bf q},{\bf k},{\bf k_{1}})
~\phi(k,t)~\phi(k_{1},t) \eea
where the vertex function $V$ is,
\be V({\bf q},{\bf k},{\bf k_{1}})=S(k)S(k_{1})~[(\hat{\bf q}.{\bf
k})~ c(k)+(\hat{\bf q}.{\bf k}_{1})~ c(k_{1})]^{2}~, \ee
using the notation ${\bf k_1}={\bf q}-{\bf k}$. $c(k)$ is the
fourier transform of Ornstein-Zernike direct correlation function
$c(r)$ introduced in the expression (\ref{fre2}) for the free
energy. $S(k)$ is the static structure factor of the liquid. Taking
$F_U$ as a quadratic functional of the fields $\rho$, the eqn. of
motion (\ref{momt-eqn}) for the momentum density ${\bf g}$ gives
rise to a cubic nonlinearity. The form of the vertex functions in
the nonlinear terms of the momentum density (${\bf g}$) equation is
linked to the both the ideal gas and excess part of the free energy
functional $F[\rho]$ in Eqn. (\ref{fre2}) presented above.

\subsection{Single particle dynamics}

\begin{figure}[h]
\includegraphics[width=0.5\textwidth]{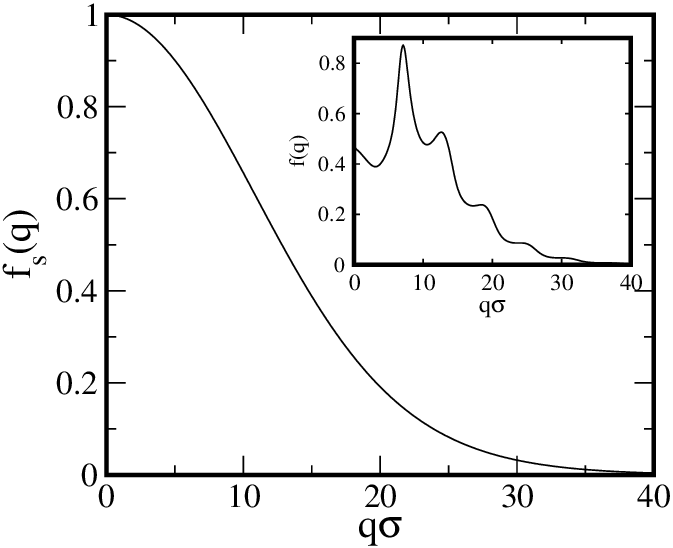}
\caption{Non ergodicity parameters $f_s$ for self and $f$ for total
correlation functions shown as the main and inset at $\eta=.525$.}
\label{fig01}
\end{figure}

At the microscopic level the density of a tagged single particle in
the fluid is a conserved property similar to the collective
densities. The Laplace transform of the normalized correlation
$\phi_s(q,t)$ is obtained in a form similar to that for $\phi(q,z)$:
\bea \lbl{self} \phi_s(q,z)=\frac{1}{z+iq^2D_s(q,z)} \eea
with the inverse of the renormalized self diffusion coefficient
\bea D_{s}^{-1}(q,z)&=&\nu_0^{-1}+\Delta_{mc}(q,z) \eea
The bare diffusion coefficient is $\nu_0=\tau_0/(\beta{m})$ where
$\tau_{0}$ presents a characteristic short time for the liquid
dynamics \cite{das-dufty,lutsk-sh}. For the supercooled dense liquid
the above memory function produces sharp fall in the tagged particle
diffusion if we make the so called adiabatic approximation
\cite{kawasaki,bin2}. The involves assuming that the density
fluctuations decays much more slowly than the momentum fluctuations
in the deeply supercooled state. With this approximation applied to
the momentum density equation the current correlation functions are
simply expressed \cite{bin2} in terms of density correlation
functions. We obtain, at one-loop order, for the mode coupling part
of the memory function for the inverse of $D_s(q,z)$
\be \label{nu4} \Delta_{mc}(q,t) =\frac{1}{\rho_0} \int\frac{d {\bf
k}}{(2\pi)^3} {\left [ (\hat{\bf q}.{\bf
k})\frac{c(k)}{\tilde{\Gamma}_0(k)}\right ]}^2 ~{\phi}_s(k_1,t)~
\phi(k,t)~S(k).~ \ee
$\Delta_{mc}$ is dependent on the vertex function involving the wave
vector dependent factor $\tilde\Gamma_0(k)$. The latter determines
the variation of the sound attenuation or damping of density
fluctuations over a range of length scales ( $k$-values) starting
from short distances. The correlation of the tagged-particle
momentum also follows in a straightforward manner from the above
analysis. The renormalized expression for the longitudinal
tagged-particle current correlation is obtained as
\be \label{self-cur} F_s(q,z)=\frac{1}{z+\tau_{0}^{-1}+v_0^2
\Delta_{mc}(q,z)} \ee
For the correlation of the single particle density, the
corresponding long time limit of the correlation function is defined
as
\be \phi_s(t\rightarrow \infty) = f_s(q)~~~~. \ee
$f_s(q)$ is referred to as the non-ergodicity parameter of self
correlation function. Within the adiabatic approximation of fast
relaxation of momentum fluctuations compared to that of density
fluctuations, both $f(q)$ and $f_s(q)$ simultaneously become nonzero
at the ENE transition. This is understood as follows: At the ENE
transition the long time limit of $\phi(t)$ is nonzero {\em i.e.},
$\phi(z)\sim{1/z}$ and the corresponding Laplace transform of the
generalized transport coefficient $L^{mc}(z)\sim{1/z}$ pole
conforming to the physics of diverging the viscosity. Using the form
(\ref{nu4}) of the memory function obtained in the adiabatic
approximation, it follows that Laplace transform of the
corresponding memory function also has the same pole {\em i.e.},
$\Delta_{mc}(z)\sim{1/z}$ . Therefore the diffusion coefficient goes
to zero at the ENE transition and beyond the ENE transition,
$\phi_s(z)\sim{1/z}$ implying that $f_s(q)\ne 0$. Taking the long
time limit of Eq.(\ref{corre})and (\ref{self}) we get a set of self
consistent integral equations for the respective non ergodicity
parameters $\{f\}$:
\be \label{NEP1} \frac{f(q)}{1-f(q)}={\cal L}(q)~~. \ee
Here ${\cal L}(q)$ is the long time limits of the renormalized
(longitudinal) viscosity $L^{mc}(q,t)/c_0^2$. The Percus-Yevick
structure factor (with verlet-Weiss correction) is used as an input
in solving the NEP equation (\ref{NEP1}) and (\ref{NEP2}). The inset
of Fig.\ref{fig01} displays the solution of the NEP equations for a
hard sphere system at packing fraction $\eta=.525$. Using the values
of the long time limits for the collective density correlations the
corresponding single particle quantity $f_s$ is obtained from the
solution of the equation
\be \label{NEP2} \frac{f_s(q)}{1-f_s(q)}=q^{-2}{\gamma(q)}~~. \ee
The functions $\gamma(q)$ is the long time limits of the
renormalized memory function $\Delta_{mc}(q,t)$ scaled with the bare
diffusion constant $\nu_0=\gamma_0\chi_{cc}^{-1}$. Since the
$\gamma(q)$ is finite in the $q{\rightarrow}0$ limit, from Eqn.
(\ref{NEP2}) it is clear that $f_s(q)$ remains pinned at the value
$1$. The $f_s(q)$ obtained using the $f(q)$ at density $\eta=.525$
is shown in Fig. \ref{fig01}.

\section{Linking DFT with MCT}
\label{sec3}

We now demonstrate that the results from the mode coupling models
for the dynamics of the frozen supercooled liquid are linked to the
static density functional description of the amorphous solid state.
This link is established here focusing on the the tagged particle
dynamics as predicted from the two different descriptions of the
frozen solid. In the thermodynamic picture, the glassy state is
characterized as one in which the individual particles are
oscillating around a set of lattice sites $\{ {\bf R}_i\}$ situated
on a random lattice. On the other hand, in the simplest form of mode
coupling theory which predict the sharp ENE transition, the self
diffusion coefficient $D_{s}(q,z=0)$ of the tagged particle becomes
zero below the dynamic transition point. This also implies complete
localization similar to the static problem. The key quantity for
analyzing the single particle dynamics is the tagged particle
density correlation function given by (\ref{self}). The mode
coupling effects are included here in terms of the self energy
$\Delta_{mc}(0,t)$ given by eqn.(\ref{nu4}). The form (\ref{nu4}) of
the mode coupling contribution to the memory function $\Delta_{mc}
(0,t)$ is obtained in the adiabatic approximation. In MCT the
transition of the supercooled liquid to the nonergodic glassy state
occurs at a critical density. This transition is driven by a key
feed back mechanism as a result of which the density correlation
function freezes in the long time limit. Freezing of the density
correlation function $\phi(q,t)$ thus implies that the Laplace
transform of the memory function $\Delta_{mc}(z) $develops a $1/z$
pole. In the small $z$ limit we write for the memory function,
\begin{equation}
\label{sp-dyn-slf}
 \Delta_{mc}(z) = \frac{\tilde\gamma_\infty}{z} + R.T
\end{equation}
where $\tilde\gamma_\infty$ is the long time limit of
$\Delta_{mc}(0,t)$,
\begin{equation}
\label{dnep-gm} \tilde\gamma_\infty =
\Delta_{mc}(0,t\rightarrow{\infty})
\end{equation}
and R.T represents the regular terms. The development of the $1/z$
pole or non zero value for $\tilde\gamma_\infty$ is therefore
crucial for the localized motion of the single particle signifying
vibrating modes. Assuming the latter to be harmonic in nature a
simple relation is reached between the corresponding spring constant
and $\tilde\gamma_\infty$. The form (\ref{sp-dyn-slf}) for the
memory function, it follows from the denominator of the RHS of
Eqn.(\ref{self-cur}) that the current correlations represent damped
harmonic waves having pole structure
${(z^2-v_0^2\gamma_\infty)}^{-1}$. The latter represents vibration
around the lattice sites. The corresponding frequency of oscillation
$\omega_s$ and hence the spring constant $\kappa_s$ of the vibration
is obtained from the pole as
\begin{equation}
\label{sp-slf-fr} \omega_0^2 =v_0^2\tilde\gamma_{\infty}~~.
\end{equation}
The above result for the frequency of the oscillator follows from a
purely dynamical route ( MCT) for the localized single particle
dynamics. Now using the relation (\ref{alreln2}) the width parameter
$\alpha$ obtained from the dynamical consideration, is related to
the self energy $\tilde\gamma_\infty$ as,
\begin{equation}
\label{mct-l2} \alpha_\mathrm{MCT} =
\frac{\tilde\gamma_\infty}{2}~~.
\end{equation}
The quantity $\tilde\gamma_\infty$ is obtained by taking the long
time limit for the memory kernel $\Delta_{mc}(t)$. It follows from
the result (\ref{nu4}):
\be \label{mem-se1} \Delta_{mc}(0,t)=\frac{1}{3\rho_0} \int\frac{d
{\bf k}}{(2\pi)^3} {\left
[k~\frac{c(k)}{\tilde{\Gamma}_0(k)}\right]}^2 ~{\phi}_s(-k,t)~
\phi(k,t)~S(k).~ \ee
The above relation then serves as the key link between the DFT
description of the harmonic solid and the mode coupling kernel. In
the ergodic state $\tilde\gamma_\infty{\rightarrow}0$ and hence
$\alpha$ vanishes corresponding to the liquid state. The equation
(\ref{mct-l2}) for $\alpha$ is now obtained in the form
\begin{equation}
\label{alp_MCT} \alpha_\mathrm{MCT} = \frac{1}{6 \rho_0}\int \frac{d
{\bf k}}{(2\pi)^3}~
\left[k~\frac{c(k)}{\tilde{\Gamma}_0(k)}\right]^2~f_s(k)~f(k)~S(k)
\end{equation}
Having discussed the two models we are now able to test the
equivalence of the two models by comparing the numerical results for
$\alpha_\mathrm{DFT}$ and $\alpha_\mathrm{MCT}$ obtained for a
simple hard sphere system using the DFT and the MCT respectively.

\section{Numerical results} \label{sec:Bernal-gb}

\begin{figure}[h]
\includegraphics[width=0.5\textwidth]{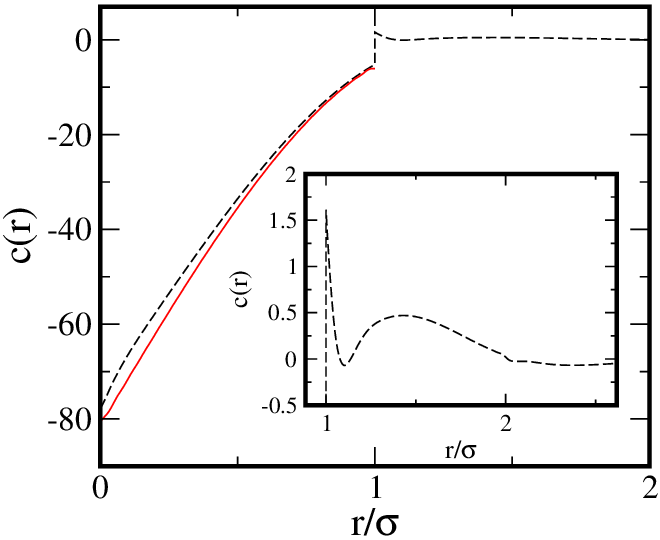}
\caption{The direct correlation function $c(r)$ with Verlet-Weiss
correction (solid) and without(dashed) for the uniform liquid at
packing $\eta=.524$. Inset is an enlarged view} \label{fig02}
\end{figure}

\begin{figure}[h]
\includegraphics[width=0.5\textwidth]{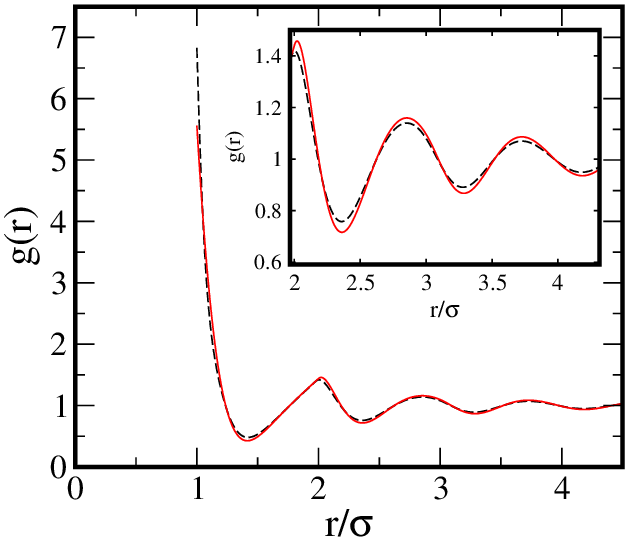}
\caption{The pair correlation function $c(r)$ with Verlet-Weiss
correction (solid) and without(dashed) for the uniform liquid at
packing $\eta=.524$. Inset is an enlarged view} \label{fig03}
\end{figure}

We use the density functional model to compute the
$\alpha_\mathrm{DFT}$ which sets the extent of mass localization in
the metastable state. The inhomogeneous density defined in Eqn.
(\ref{tar-dens}) is chosen to evaluate the free energy functional
(\ref{fre2}). The density distribution in a highly inhomogeneous
state of the liquid is represented by sharply peaked Gaussian
functions centered around each of the lattice sites. The locations
of the centers for the Gaussian density profiles, $\{{\bf R}_i\}$
distributed on a random lattice, is an input required for the
density function. Assuming that the Bernal's random structure
\cite{Bernal} is a good approximation for describing the supercooled
liquid structure, we use the site-site correlation function $g(R)$
of the corresponding random structure. This pair correlation is
determined using the Bennett's algorithm \cite{Benett} and has been
used in earlier DFT studies \cite{ysingh,lowen} for describing
typical glassy structures. Following Baus and Colot
\cite{baus-colot}, we use the random structure $g_s(R)$ through the
following relation
\begin{equation} \label{bern-en}
g_s(R)=g_B[ R {(\eta/\eta_0)}^{1/3}],
\end{equation}
where $\eta$ denotes the average packing fraction and $\eta_0$ is
used as a scaling parameter for the structure  such that at
$\eta=\eta_0$ Bernal's structure $g_B(R)$ is reproduced. $g_B({\bf
R})$ is computed by averaging over the site-site correlations
obtained by taking different points as origin throughout the random
structure. The mapping of the function from ${\bar R}=R
(\eta/\eta_0)^{1/3}$ to $R$ allows the structure to either dilute or
contract depending on the scaling parameter $\eta_0$. The set of
lattice sites $\{{\bf R}_i\}$ found in computer simulation studies
for average positions of particles has also been used in searching
the free energy minimum \cite{kim-munakata}. Similar metastable
minima of the free energy functional as above has been  observed
from such studies as well. In this asymptotic limit corresponding
to large $\alpha$, the ideal gas part of free energy
(Eqn.(\ref{fid_ge})) reduces to,
\be \label{fid_app} \beta \Delta f_{id}[\rho] \approx -\left [
\frac{3}{2} + \ln~ \rho_0 - \frac{3}{2}~\ln \left(
\frac{\pi}{\alpha} \right) \right ]. \ee
The above formula is obtained by replacing the summation inside the
logarithmic term by just the contribution from the nearest site.
This reduction is not valid for the weakly localized state of the
amorphous structure where the $\alpha$ values are small. In this
case the gaussian functions centered at different lattice sites are
wide enough to cause considerable overlap and the above reduction
can no longer be justified. Thus in the range of small $\alpha$ we
numerically evaluate the equation (\ref{fid_ge}) and the asymptotic
form given in Eqn. (\ref{fid_app}) is used in the large $\alpha$
regions. In our earlier work \cite{prl1} we have shown that these
two expressions start merging for typically $\alpha>40$ within
$1\%$. Similarly we also evaluate the excess contribution to the
free energy by expanding around the uniform liquid state. The excess
contribution to the free energy therefore depends on the structure
of the uniform liquid which is described in terms of the direct
correlation function $c(r)$. For the $c(r)$, we have used the
Percus-Yevick (PY) form with Verlet-Weis correction as obtained by
Henderson and Grundke \cite{VW,HG}. We display in Fig. \ref{fig02}
the direct correlation function $c(r)$ at $\eta=.524$ with and
without the Verlet-Weiss correction. These results for the uniform
liquid structure is used for both DFT as well as the MCT based
calculations which are to be presented in the following. We also
display the corresponding results for $g(r)$ for $r/\sigma$ between
2 and 4, in the inset of Fig. \ref{fig03}.

%%%%%%%%%%%%

\begin{figure}[h]
\includegraphics[width=0.5\textwidth]{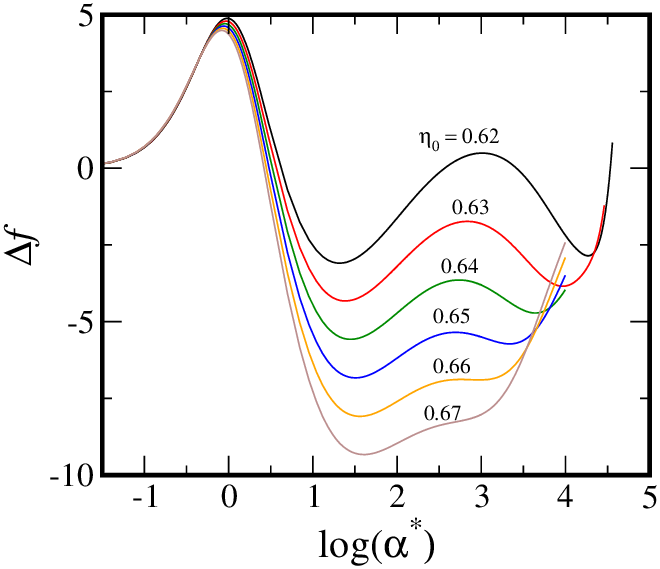}
\caption{Free energy vs. $\alpha$ at fixed $\eta=.597$ for different
structures characterized in terms of the parameter $\eta_0$ (see
text) as indicated with the corresponding curve.} \label{fig04}
\end{figure}

\begin{figure}[h]
\includegraphics[width=0.5\textwidth]{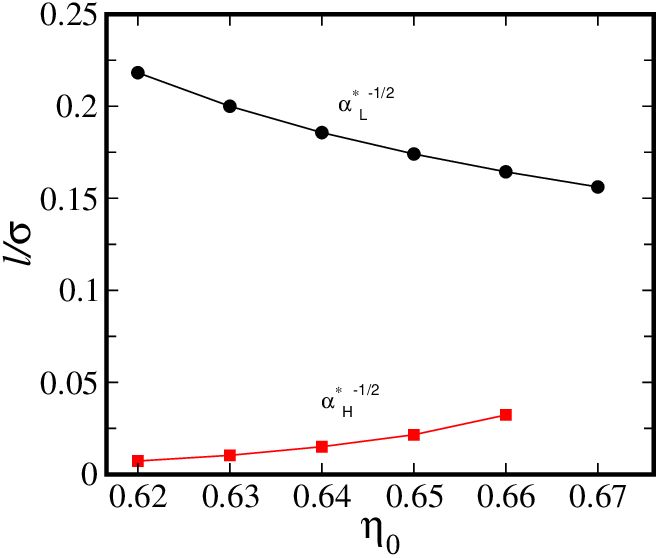}
\caption{The average mean square displacement $\ell$ in units of
hard sphere diameter $\sigma$, for both high localization and low
localization vs $\eta_0$ (see text), corresponding to the packing
fraction $\eta=.597$.} \label{fig05}
\end{figure}

The free energy functional $F[\rho]$ is computed for different
choices of the test density function $\rho({\bf r})$ which is
dependent on the localization parameter $\alpha$. We consider how
the free energy surface changes at a fixed packing $\eta$ for
different input structures characterized by the inhomogeneous
density $\rho({\bf r})$. For a fixed packing fraction, we calculate
the $F[\rho]$ as a function of the parameter $\eta_0$. There are two
minimum of the free energy at two respective values of the width
parameter $\alpha$. The first minimum corresponding to a relatively
smaller value of $\alpha$ represents a metastable state with low
degree of mass localization compared to the second minimum which
appears at a larger $\alpha\sigma^2$ value. Different choices of
$\eta_0$ in the pair correlation function $g_s(R)$, defined in Eqn.
(\ref{bern-en}), are equivalent to choosing different underlying
structures for the amorphous lattice depicting the centers of the
gaussian density profiles. Testing the free energy dependence on
$\alpha$ at a fixed $\eta$, we find that the two minima form
persists only over a range of $\eta_0$. The free energy curves for
different $\eta_0$ corresponding to a fixed packing fraction
$\eta=.597$ is shown in Fig. \ref{fig04}. The optimum $\alpha$ at
the minimum of the free energy is termed as $\alpha_\mathrm{DFT}$.
The low and high $\alpha_\mathrm{DFT}$ minima are referred to here
as $\alpha_\mathrm{L}$ and $\alpha_\mathrm{H}$. The mean square
displacement is defined as $\ell=1/\sqrt{\alpha_\mathrm{DFT}}$. The
dependence of the mean square displacement $\ell$ on $\eta_0$ (at a
fixed value of $\eta$) is displayed in Fig. \ref{fig05}.

%%%%%%%%%%%%%%%

\begin{figure}[h]
\includegraphics[width=0.5\textwidth]{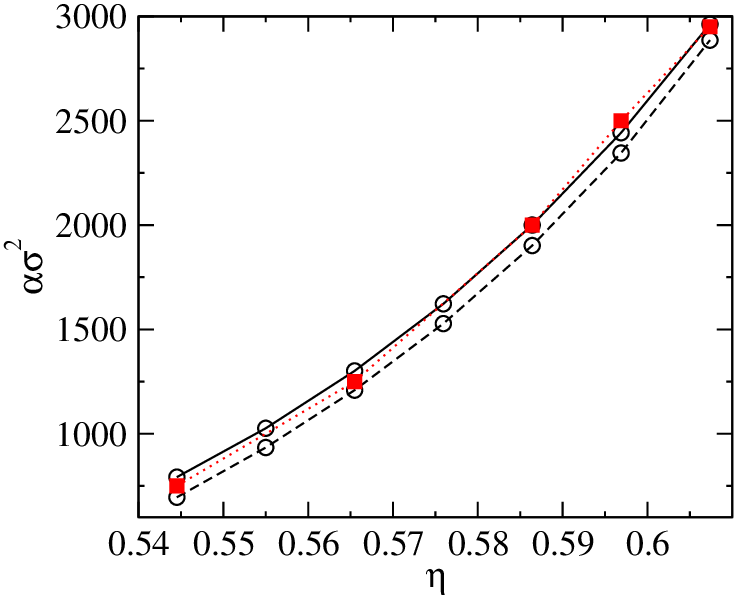}
\caption{Comparison of the width parameter values (in units of
$\sigma^{-2}$) from DFT (filled squares) and MCT (open circles).
$\alpha_\mathrm{MCT}$ obtained from MCT using as input Percus Yevick
structure factors with  (dashed) and with out (solid) the
Verlet-Weiss correction. The result $\alpha_\mathrm{DFT}$ obtained
from DFT is shown as dotted line.} \label{fig06}
\end{figure}

\begin{figure}[h]
\includegraphics[width=0.5\textwidth]{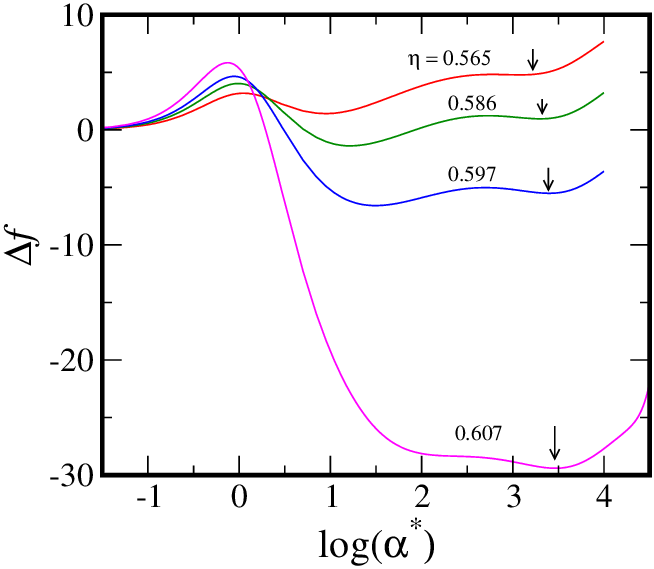}
\caption{Free energy vs. $\alpha$ for different packing fractions
$\eta$ denoted in Fig. \ref{fig06}. The position of the
corresponding minima at large $\alpha$ is shown with a vertical
arrow for each free energy curve}\label{fig07}
\end{figure}

\begin{figure}[h]
\includegraphics[width=0.5\textwidth]{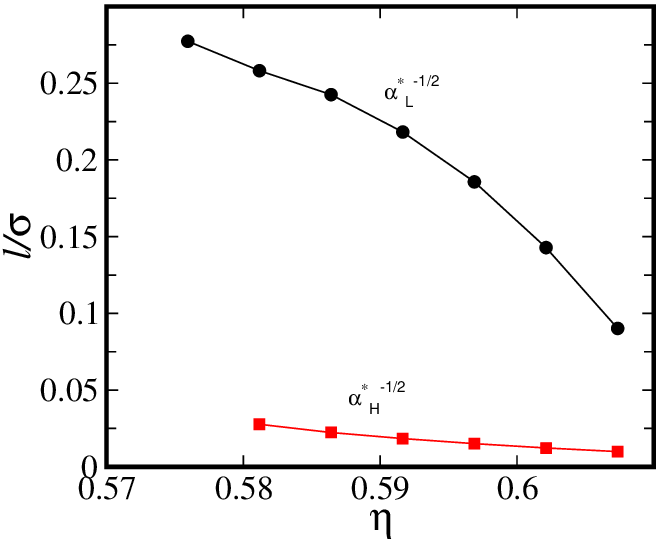}
\caption{The average mean square displacement $\ell$ in units of
hard sphere diameter $\sigma$, for both high localization and low
localization vs packing fraction $\eta$, corresponding to the
results shown in Fig. \ref{fig06}.} \label{fig08}
\end{figure}

\begin{figure}[h]
\includegraphics[width=0.5\textwidth]{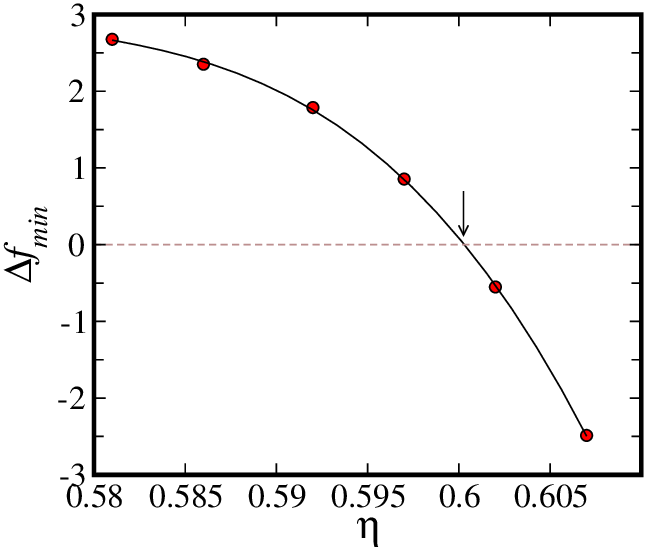}
\caption{The difference of the free energy minima at low and high
values of $\alpha$ as shown in Fig. \ref{fig06} vs packing fraction
$\eta$ corresponding to the results shown in \ref{fig06}. }
\label{fig09}
\end{figure}

\begin{figure}[h]
\includegraphics[width=0.5\textwidth]{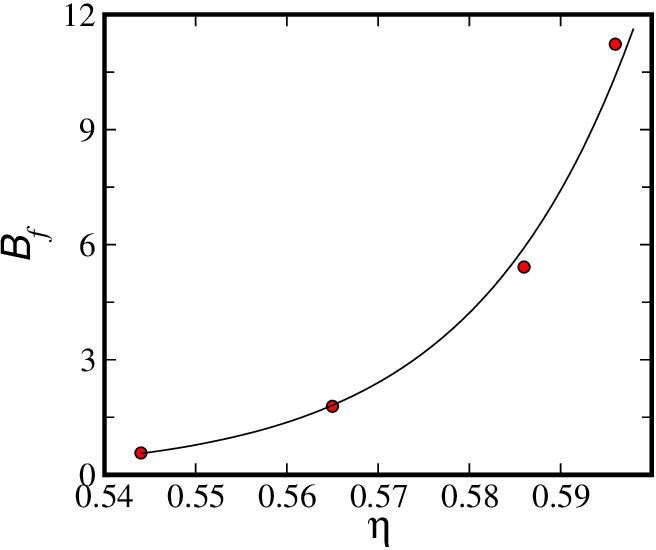}
\caption{The barrier in free energy ${\cal B}_f$  measured from the
low $\alpha$ minimum, to cross to the uniform liquid state at
$\alpha=0$ vs. packing fraction $\alpha$. The line shows a power law
fitting ${(\eta_c-\eta)}^{-a}$} \label{fig10}
\end{figure}

Having seen the basic relation between the free energy surface and
the localization parameter $\alpha$ we set to compare the MCT and
DFT approaches to single particle dynamics in the supercooled state.
With increased packing the free energy minimum with high $\alpha$
signifying the sharply localized particle configurations get more
stable. We compare between a) the mass localization parameter
$\alpha_\mathrm{DFT}$ calculated using the DFT and b) the
corresponding value of $\alpha$ obtained from Eqn. (\ref{alp_MCT})
using dynamical approach of MCT. In Fig. \ref{fig06} the results for
the localization parameters obtained using the DFT and MCT
respectively are shown. As the packing fraction increases the
density profiles are sharper, {\em i.e.}, the optimum $\alpha$
increases. For the variation of $\eta$ along the curve shown in Fig.
\ref{fig06} we change the parameter $\eta_0$ with $\eta$ so as to
keep the ratio $\gamma_0={(\eta/\eta_0)}^{1/3}$ remains fixed at
$.97$. The free energy curves for the corresponding pairs of $\eta$
and $\eta_0$ values are shown in Fig. \ref{fig07}. For each of these
respective free energy curves in Fig. \ref{fig07}, the corresponding
MCT value of $\alpha$ at the same packing fraction, is indicated on
the $\alpha$-axis with a vertical arrow.
The position of the arrow depicts the corresponding
$\alpha_\mathrm{MCT}$ which is calculated using model of completely
localized single particle dynamics. It appears to be close to the
corresponding $\alpha_\mathrm{DFT}$ for the metastable minimum
having the {\em sharply} localized density profiles. Fig.
\ref{fig08} shows how the localization lengths $\ell$ for the
particles corresponding to the  high and low  $\alpha_\mathrm{DFT}$
metastable states depends on packing fraction $\eta$. The less
localized minimum (corresponding to small $\alpha$ values) is more
robust below packing fraction $.60$. With increasing packing
fraction, the sharply localized state (corresponding to large
$\alpha$ values) becomes more stable. In Fig. \ref{fig09} we show
(with an arrow at $\eta\approx{.60}$) the cross over in the relative
stability by plotting the difference of the two free energies with
respect to packing fraction $\eta$. In Fig. \ref{fig10} we display
how the height ${\cal B}_f$ of the barrier to cross over from the
metastable to state at finite $\alpha$ to the uniform liquid state (
at $\alpha\rightarrow{0}$) grows with increase of packing fraction.

\section{Discussion}

For the single particle in the supercooled state, the average
potential energy of the oscillator is given by $\kappa{x^2}/2$.
Application of the equipartition law obtains a relation between the
spring constant $\kappa$ and the width parameter $\alpha$. This
estimation is therefore based on the basis of a static harmonic
potential for the single particle dynamics. The same spring constant
and hence the natural frequency of the oscillator is also obtained
using the MCT approach. The simplest form of MCT predicts an
ergodic-nonergodic transition in the supercooled liquid. Beyond the
transition, in the non-ergodic state, the single particle
correlation decays like a harmonic wave. The frequency of the wave
is related to the long time limit of the memory function
\cite{boon-yip} corresponding to the correlation of single particle
densities. This links natural frequency of oscillation and hence the
memory function is related to the spring constant $\kappa$. The
width parameter $\alpha$ of DFT is thus related to the long time
limit of the memory function.

%%%%%%%%%% Role of Free energy functional

Phenomenological relation between structural properties and
transport coefficients of a liquid has earlier been observed
\cite{dgutov,entr-jcp}. In the present work, we find agreement
between the static and dynamic models of the metastable liquid in
which the particle positions are localized over time scales of
structural relaxation. This is concluded from the close agreement
obtained for the width parameter $\alpha$ in the respective
theoretical models of the DFT and MCT. The primary connection
between the DFT and MCT formulations is the free energy functional
$F[\rho]$ which appears in the formulation of either model. The free
energy determines both the structure as well as the dynamics. In the
thermodynamic approach the local minima of $F[\rho]$  represent the
metastable states. In the dynamic approach the same free energy
functional is key to the form of the nonlinear couplings in the
equations of generalized hydrodynamics. These nonlinearities are the
essential ingredient in construction of the memory functions for the
dynamics. The self consistent expression for the memory functions
constitute the MCT with the feedback mechanism that causes the ENE
transition at a critical density. In the nonergodic state the single
particle motion is localized. Hence in our comparison, using the
same free energy functional in constructing both the DFT and the MCT
models is essential. However the approximate free energy functional
(\ref{fre2}) is not good when the density fluctuations are large,
which is more the case near freezing point $T_m$. With the strongly
localized density profiles of the amorphous state, this same concern
remains. In this regard the DFT has been improved using weighted
density functional \cite{denton,spd-pre} models. In such models the
localization parameter changes significantly. However to compare
with the MCT results would then also require changing the mode
coupling vertex functions in the dynamic model accordingly.
Furthermore, in the present work the density profiles are assumed to
be described by gaussian profiles and all non-gaussian behavior are
ignored \cite{prl_phonon}.

%%%% The key observation - low and high alpha minima

From the density functional approach it emerges that the free energy
functional expressed in terms of density fluctuations, have two
qualitatively different types of minimum \cite{prl1}. The minimum of
the $F[\rho]$ corresponding to a lower degree of mass localization
(at lower $\alpha$ value) is generally more stable than the state
with sharply localized density profiles (higher $\alpha$ values).
Both minima (low and high $\alpha$ values) represents particle
localization and hence conforms to a nonergodic state. However as we
note from the results of Fig. \ref{fig06}-\ref{fig07}, this less
localized minimum corresponds to a higher mean square displacement
for the tagged particle than what is predicted for the nonergodic
state obtained from a simple one loop MCT for self correlations
\cite{bin2}. This can possibly be an outcome of the simple adiabatic
approximation we have adopted \cite{bin2} in evaluating the
corresponding memory function for the single particle dynamics. Over
intermediate time scales however, the interpretation of sharp
localization given in Ref. \cite{dftmct} for the high $\alpha$
minimum and hence agreement the simple MCT models holds. This
similarity between the two microscopic models for describing the
physics of supercooled liquid is a signature of the consistency in
the respective approached. Over asymptotically long time scales the
particle should gets out of the non ergodic state \cite{DM09,pre96}.
We have not considered here the mechanisms which might led to
spontaneous breaking of ergodicity in the supercooled liquid. With
this an exponentially large number of metastable states appear for
intermediate free energy values and gives rise to finite
configurational entropy \cite{monasson} for the system.

\section*{Acknowledgement}
\label{sec_ACK} LP and NB acknowledges CSIR, India for financial
support. SPD acknowledges financial support under UPOE grant of
Jawaharlal Nehru University, New Delhi.

\end{document}